# Complexity of Mechanism Design

**Vincent Conitzer** and **Tuomas Sandholm**
{conitzer, sandholm}@cs.cmu.edu
Computer Science Department
Carnegie Mellon University
Pittsburgh PA 15211

## Abstract

The aggregation of conflicting preferences is a central problem in multiagent systems. The key difficulty is that the agents may report their preferences insincerely. *Mechanism design* is the art of designing the rules of the game so that the agents are motivated to report their preferences truthfully and a (socially) desirable outcome is chosen. We propose an approach where a mechanism is automatically created for the preference aggregation setting at hand. This has several advantages, but the downside is that the mechanism design optimization problem needs to be solved anew each time. Focusing on settings where side payments are not possible, we show that the mechanism design problem is $\mathcal{N}P$-complete for deterministic mechanisms. This holds both for dominant-strategy implementation and for Bayes-Nash implementation. We then show that if we allow randomized mechanisms, the mechanism design problem becomes tractable. In other words, the coordinator can tackle the computational complexity introduced by its uncertainty about the agents' preferences by making the agents face additional uncertainty. This comes at no loss, and in some cases at a gain, in the (social) objective.

## 1 Introduction

In multiagent settings, agents generally have different preferences, and it is of central importance to be able to aggregate these, i.e. to pick a socially desirable *outcome* from a set of outcomes. Such outcomes could be potential presidents, joint plans, allocations of goods or resources, etc. For example, voting mechanisms constitute an important class of preference aggregation methods.

The key problem is the following uncertainty. The coordinator of the preference aggregation generally does not know the agents' preferences *a priori*. Rather, the agents report their preferences to the coordinator. Unfortunately, in this setting, most naive preference aggregation mechanisms suffer from *manipulability*. An agent may have an incentive to misreport its preferences in order to mislead the mechanism into selecting an outcome that is more desirable to the agent than the outcome that would be selected if the agent revealed its preferences truthfully. Manipulation is an undesirable phenomenon because preference aggregation mechanisms are tailored to aggregate preferences in a socially desirable way, and if the agents reveal their preferences insincerely, a socially undesirable outcome may be chosen.

Manipulability is a pervasive problem across preference aggregation mechanisms. A seminal negative result, the *Gibbard-Satterthwaite theorem*, shows that under *any* nondictatorial preference aggregation scheme, if there are at least 3 possible outcomes, there are preferences under which an agent is better off reporting untruthfully [9, 19]. (A preference aggregation scheme is called dictatorial if one of the agents dictates the outcome no matter how the others vote.)

In the case of software agents, the algorithms they use for deciding how to report their preferences must be coded explicitly. Given that the reporting algorithm needs to be designed only once (by an expert), and can be copied to large numbers of agents (even ones representing unsophisticated humans), it is likely that manipulative preference reporting will increasingly become an issue, unmuddied by irrationality, emotions, etc.

What the coordinator would like to do is to design a preference aggregation mechanism so that 1) the self-interested agents are motivated to report their preferences truthfully, and 2) the mechanism chooses an outcome that is desirable from the perspective of some social objective. This is the classic setting of *mechanism design* in game theory. In mechanism design, there are two different types of uncertainty: the coordinator's uncertainty about the agents' preferences, and the agents' uncertainty about each others' preferences (which can affect how each agent tries to manipulate the mechanism).

Mechanism design provides us with a variety of carefully crafted definitions of what it means for a mechanism to be nonmanipulable, and goals to pursue under this constraint (for example, maximization of social welfare). It also provides us with some general mechanisms which, under certain assumptions, are nonmanipulable and socially desirable (among other properties). The upside of these mechanisms is that they do not rely on (even probabilistic) information about the agents' preferences (e.g., the Vickrey-Groves-Clarke mechanism [5, 10, 20]), or they can be easily applied to any probability distribution over the preferences (e.g., the dAGVA mechanism [1, 7]). The downside is that these mechanisms only work under very restrictive assumptions, the most common of which is to assume that the agents can make side payments and have quasilinear utility functions. The quasilinearity assumption is quite unrealistic. It means that each agent is risk neutral, does not care about what happens to its friends or enemies, and values money independently of all other attributes of the outcome. Also, in many voting settings, the use of side payments would not be politically feasible. Furthermore, among software agents, it might be more desirable to construct mechanisms that do not rely on the ability to make payments.

In this paper, we propose that the *mechanism be designed automatically for the specific preference aggregation problem at hand.* We formulate the mechanism design problem as an optimization problem. The input is characterized by the number of agents, the agents' possible types (preferences), and the coordinator's prior probability distributions over the agents' types. The output is a nonmanipulable mechanism that is optimal with respect to some (social) objective.

This approach has three advantages over the classic approach of designing general mechanisms. First, it can be used even in settings that do not satisfy the assumptions of the classical mechanisms. Second, it may yield better mechanisms (in terms of stronger nonmanipulability guarantees and/or better social outcomes) than the classical mechanisms. Third, it may allow one to circumvent impossibility results (such as the Gibbard-Satterthwaite theorem) which state that there is no mechanism that is desirable across all preferences. When the mechanism is designed to the setting at hand, it does not matter that it would not work on preferences beyond those in that setting.

However, this approach requires the mechanism design optimization problem to be solved anew for each preference aggregation setting. In this paper we study how hard this computational problem is under the two most common nonmanipulability requirements: dominant strategies, and Bayes-Nash equilibrium [16]. We will study preference aggregation settings where side payments are not an option.

The rest of the paper is organized as follows. In Section 3 we define the *deterministic* mechanism design problem, where the mechanism coordinator is constrained to deterministically choose an outcome on the basis of the preferences reported by the agents. We then show that this problem is $\mathcal{NP}$-complete for the two most common concepts of nonmanipulability. In Section 4 we generalize this to the *randomized* mechanism design problem, where the mechanism coordinator may stochastically choose an outcome on the basis of the preferences reported by the agents. (On the side, we demonstrate quickly that reandomized mechanisms may be strictly more efficient than deterministic ones.) We then show that this problem is in fact solvable by linear programming for both concepts of nonmanipulability.

## 2 The setting

Before we define the computational problem of mechanism design, we should justify our focus on nonmanipulable mechanisms. After all, it is not immediately obvious that there are no manipulable mechanisms that, even when agents report their types strategically and hence sometimes untruthfully, still reach better outcomes (according to whichever objective we use) than any nonmanipulable mechanism. Additionally, given our computational focus, we should also be concerned that manipulable mechanisms that do as well as nonmanipulable ones may be easier to find. It turns out that we need not worry about either of these points: given any mechanism, we can quickly construct a nonmanipulable mechanism whose performance is exactly identical. For given such a mechanism, we can build an interface layer between the agent and this mechanism. The agents input (some report of) their preferences (or *types*) into the interface layer; subsequently, the interface layer inputs the types *that the agents would have strategically reported* if their types were as declared into the original mechanism, and the resulting outcome is the outcome of the new mechanism. Since the interface layer acts "strategically on each agent's best behalf", there is never an incentive to report falsely to the interface layer; and hence, the types reported by the interface layer are the strategic types that would have been reported without the interface layer, so the results are exactly as they would have been with the original mechanism. This argument (or at least the existential part of it, if not the constructive) is known in the mechanism design literature as the *revelation principle* [16]. Given this, we can focus on truthful mechanisms in the rest of the paper.

We first formally define a preference aggregation setting.

**Definition 1** *A* preference aggregation setting *consists of a set of outcomes $O$, a set of agents $A$ with $|A| = N$, and for each agent:*

- *A set of types $\Theta^i$;*
- *A probability distribution $p_i$ over $\Theta^i$;[1]*
- *A utility function $u_i : \Theta^i \times O \to \mathcal{Q}$;[2]*

*These are all common knowledge. However, each agent's type $\theta_i \in \Theta_i$ is private information.*

Though this follows standard game theory notation, the fact that agents have both utility functions and types is perhaps confusing. The types encode the various possible preferences that agents may turn out to have, and the agents' types are not known by the coordinator. The utility functions are common knowledge, but the agent's type is a parameter in the agent's utility function. So, the utility of agent $i$ is $u_i(\theta^i, o)$, where $o \in O$ is the outcome and $\theta^i$ is the agent's type.

## 3 Deterministic mechanisms

We now formally define a deterministic mechanism.

**Definition 2** *Given a preference aggregation setting, a* deterministic mechanism *is a function that, given any vector of reported types, produces a single outcome. That is, it is a function $o : \Theta^1 \times \Theta^2 \times \ldots \times \Theta^N \to O$.*

A *solution concept* is some definition of what it means for a mechanism to be nonmanipulable. We now present the two most common solution concepts. They are the ones analyzed in this paper.

**Definition 3** *Given a preference aggregation setting, a mechanism is said to* implement its outcome function in dominant strategies *if truthtelling is always optimal, no matter what types the other agents report. For a deterministic mechanism, this means that for any $i \in A$, for any type vector $(\theta^1, \theta^2, \ldots, \theta^i, \ldots, \theta^N) \in \Theta^1 \times \Theta^2 \times \ldots \times \Theta^i \times \ldots \times \Theta^N$, and for any $\hat{\theta}^i \in \Theta^i$, we have $u_i(\theta^i, o(\theta^1, \theta^2, \ldots, \theta^i, \ldots, \theta^N)) \geq u_i(\theta^i, o(\theta^1, \theta^2, \ldots, \hat{\theta}^i, \ldots, \theta^N))$.*

---

[1]This assumes that the agents' types are drawn independently. If this were not the case, the input to the mechanism design algorithm would include a *joint* probability distribution over the agents' types instead. All of the results of this paper apply to that setting as well.

[2]We restrict our attention to rational numbers instead or real numbers so that we do not have to worry about the fact that a real number in the input could take an infinite number of digits to represent (thus potentially allowing cheating in the analysis because exponential computations could be conducted in polynomial time in the size of the input).

Dominant strategy implementation is very robust. It does not rely on the prior probability distributions being (commonly) known, or the types being private information. Furthermore, each agent is motivated to report truthfully even if the other agents report untruthfully, for example due to irrationality.

The second most prevalent solution concept in mechanism design, *Bayes-Nash equilibrium*, does not have any of these robustness benefits. Often there is a trade-off between the robustness of the solution concept and the social welfare (or some other goal) that we expect to attain, so both concepts are worth investigating.

**Definition 4** *Given a preference aggregation setting, a mechanism is said to* implement its outcome function in Bayes-Nash equilibrium *if truthtelling is always optimal as long as the other agents report truthfully. For a deterministic mechanism, this means that for any $i \in A$, and for any $\theta^i, \hat{\theta}^i \in \Theta^i$, we have $E_{\theta^{-i}}(u_i(\theta^i, o(\theta^1, \theta^2, \ldots, \theta^i, \ldots, \theta^N))) \geq E_{\theta^{-i}}(u_i(\theta^i, o(\theta^1, \theta^2, \ldots, \theta^i_0, \ldots, \theta^N)))$. (Here $E_{\theta^{-i}}$ means the expectation taken over the types of all agents except $i$, according to their true type distributions $p_j$.)*

We are now ready to define the computational problem that we study.

**Definition 5 (DETERMINISTIC-MECHANISM-DESIGN)** *We are given a preference aggregation setting, a solution concept, and an objective function $g : \Theta^1 \times \Theta^2 \times \ldots \times \Theta^n \times O \to \mathcal{Q}$ with a goal $G$. We are asked whether there exists a deterministic mechanism for the preference aggregation setting which*

- *satisfies the given solution concept, and*
- *attains the goal, i.e. $E(g(\theta^1, \theta^2, \ldots, \theta^N, o(\theta^1, \theta^2, \ldots, \theta^N))) \geq G$. (Here the expectation is taken over the types of all the agents, according to the distributions $p_j$.)*

One common objective function is the *social welfare* function $g(\theta^1, \theta^2, \ldots, \theta^N, o) = \sum_{i \in A} u_i(\theta_i, o)$.

If the number $N$ of agents is unbounded, specifying an outcome function $o$ will require exponential space, and in this case it is perhaps not surprising that computational complexity becomes an issue even for the decision variant of the problem given here. However, we will demonstrate $\mathcal{NP}$-completeness even in the case where the number of agents is fixed at 2, and $g$ is restricted to be the social welfare function, for both solution concepts. We begin with dominant strategy implementation: for this, we will reduce from the $\mathcal{NP}$-complete INDEPENDENT-SET problem [8].

**Definition 6 (INDEPENDENT-SET)** *We are given a graph $(V, E)$ and a goal $K$. We are asked*

*whether there exists a subset $S \subseteq V$, $|S| = K$ such that for any $i, j \in S$, $(i, j) \notin E$.*

**Theorem 1** *DETERMINISTIC-MECHANISM-DESIGN with dominant strategies implementation (DMDDS) is NP-complete, even with 2 agents and the social welfare function as the objective function.*

**Proof**: To show that the problem is in $\mathcal{NP}$, we observe that specifying an outcome function here requires only the specification of $|\Theta^1| \times |\Theta^2|$ outcomes, and given such an outcome function we can verify whether it meets our requirements in polynomial time. To show NP-hardness, we reduce an arbitrary INDEPENDENT-SET instance to the following DMDDS instance. There are 2 agents, whose type sets are as follows: $\Theta_1 = \{\theta_1^1, \theta_2^1, \ldots, \theta_n^1\}$ and $\Theta_2 = \{\theta_1^2, \theta_2^2, \ldots, \theta_n^2\}$, each with a uniform distribution. For every pair $i, j$ ($1 \leq i \leq n$ and $1 \leq j \leq n$), there are the following outcomes: if $i = j$, we have $o_{ii}^H$ and $o_{ii}^L$; if $i \neq j$, and $(i, j)$ is not an edge in the graph, we have $o_{ij}$; if it is an edge in the graph, we have $o_{ij}^1$ and $o_{ij}^2$. The utility functions are as follows:

- $u_1(\theta_i^1, o_{ii}^H) = u_2(\theta_j^2, o_{jj}^H) = 2$;
- $u_1(\theta_i^1, o_{kk}^H) = u_2(\theta_j^2, o_{ll}^H) = 2$ for $i \neq k$ and $j \neq l$;
- $u_1(\theta_i^1, o_{ii}^L) = u_2(\theta_j^2, o_{jj}^L) = 1$;
- $u_1(\theta_i^1, o_{kk}^L) = u_2(\theta_j^2, o_{ll}^L) = -5n^2$ for $i \neq k$ and $j \neq l$;
- $u_1(\theta_i^1, o_{ij}) = u_2(\theta_j^2, o_{ij}) = 2$ for $i \neq j$, $(i, j) \notin E$;
- $u_1(\theta_i^1, o_{kl}) = u_2(\theta_j^2, o_{kl}) = -5n^2$ for $k \neq l$, $(k, l) \notin E$, $i \neq k$ and $j \neq l$;
- $u_1(\theta_i^1, o_{ij}^1) = u_2(\theta_j^2, o_{ij}^2) = 5n^2$ for $i \neq j$, $(i, j) \in E$;
- $u_1(\theta_i^1, o_{ij}^2) = u_2(\theta_j^2, o_{ij}^1) = 1$ for $i \neq j$, $(i, j) \in E$;
- $u_1(\theta_i^1, o_{kl}^1) = u_2(\theta_j^2, o_{kl}^1) = u_1(\theta_i^1, o_{kl}^2) = u_2(\theta_j^2, o_{kl}^2) = -5n^2$ for $k \neq l$, $(k, l) \in E$, $i \neq k$ and $j \neq l$.

Finally, we set $G = \frac{2m(5n^2+1)+2(n-K)+4K+4(n^2-2m-n)}{n^2}$. We now proceed to show that the two problem instances are equivalent.

First suppose there is a solution to the INDEPENDENT-SET instance, that is, a subset $S$ of $V$ of size $K$ such that for any $i, j \in S$, $(i, j) \notin E$. Let the outcome function be as follows:

- $o(\theta_i^1, \theta_i^2) = o_{ii}^H$ if $i \in S$;
- $o(\theta_i^1, \theta_i^2) = o_{ii}^L$ if $i \notin S$;
- $o(\theta_i^1, \theta_j^2) = o_{ij}$ if $i \neq j$, $(i, j) \notin E$;
- $o(\theta_i^1, \theta_j^2) = o_{ij}^1$ if $i \neq j$, $(i, j) \in E$, and $j \in S$;
- $o(\theta_i^1, \theta_j^2) = o_{ij}^2$ otherwise.

First we show that this mechanism is incentive compatible. We first demonstrate that agent 1 never has an incentive to misreport. Suppose agent 1's true type is $\theta_i^1$ and agent 2 is reporting $\theta_j^2$. Then agent 1 never has an incentive to instead report $\theta_k^1$ with $k \neq i, k \neq j$, since this will lead to the selection of $o_{kj}$, $o_{kj}^1$, or $o_{kj}^2$, all of which give agent 1 utility $-5n^2$. What about reporting $\theta_j^1$ instead (when $i \neq j$)? If $j \notin S$, this will lead to utility $-5n^2$ for agent 1, so in this case there is no incentive to do so. If $j \in S$, this will lead to utility 2 for agent 1. However, if $(i, j) \notin E$, reporting truthfully will also give agent 1 utility 2; and if $(i, j) \in E$, then since $j \in S$, reporting truthfully will in fact give agent 1 utility $5n^2$. It follows that again, there is no incentive to misreport.

To show that agent 2 has no incentive to misreport, first we establish that $o(\theta_i^1, \theta_j^2) = o_{ij}^2$ if $i \neq j$, $(i, j) \in E$, and $i \in S$; then, the exact same argument as for agent 1 goes through. All that is necessary to show is that in this case, $j \notin S$. But this follows immediately from the fact that there are no edges between elements in $S$, since $i \in S$ and $(i, j) \in E$. Hence, we have established incentive compatibility.

All that is left to show is that we reach the goal. Suppose the agents' types are $\theta_i^1$ and $\theta_j^2$. If $(i, j)$ is an edge, which happens with probability $\frac{2m}{n^2}$, we get a social welfare of $5n^2 + 1$. If $i = j$ and $i \notin S$, which happens with probability $\frac{n-K}{n^2}$, we get a social welfare of 2. If $i = j$ and $i \in S$, which happens with probability $\frac{K}{n^2}$, we get a social welfare of 4. Finally, if $i \neq j$ and $(i, j)$ is not an edge, which happens with probability $\frac{n^2-2m-n}{n^2}$, we get a social welfare of 4. Adding up all the terms we find that the goal is exactly satisfied.

Now suppose the mechanism design instance has a solution, that is, a function $o$ satisfying the desired properties. Suppose there are $r_1$ distinct vectors $(p, \theta_{i_1}^1, \theta_{i_2}^2)$ such that $u_p(\theta_{i_p}^p, o(\theta_{i_1}^1, \theta_{i_2}^2)) = 5n^2$ (that is, $r_1$ cases where someone gets the very large payoff) and $r_2$ distinct vectors $(p, \theta_{i_1}^1, \theta_{i_2}^2)$ such that $u_p(\theta_{i_p}^p, o(\theta_{i_1}^1, \theta_{i_2}^2)) = -5n^2$. Let $r = r_1 - r_2$. Then, since the highest other payoff that can be reached is 2, the expected social welfare can be at most $\frac{r(5n^2)+4n^2}{n^2}$. But the goal is greater than $\frac{2m(5n^2)}{n^2}$, and since by assumption $o$ reaches this goal, we have $r > 2m$. Of course, we can only achieve a payoff of $5n^2$ with outcomes $o_{ij}^1$ or $o_{ij}^2$. However, if it is the case that $o(\theta_i^1, \theta_j^2) \in \{o_{kl}^1, o_{kl}^2\}$ for $k \neq i$ or $l \neq j$, one of the agents in fact receives a utility of $-5n^2$, and the contribution of this outcome to $r$ can be at most 0. It follows that there must be at least $2m$ pairs

$(\theta_i^1, \theta_j^2)$ whose outcome is $o_{ij}^1$ or $o_{ij}^2$. But this is possible only if $(i,j)$ is an edge, and there are only $2m$ such pairs. It follows that whenever $(i,j)$ is an edge, $o_{ij}^1$ or $o_{ij}^2$ is chosen. Now, we set $S = \{i : o(\theta_i^1, \theta_i^2) = o_{ii}^H\}$. First we show that $S$ is an independent set. Suppose not: i.e. $i, j \in S$ and $(i,j) \in E$. Consider $o(\theta_i^1, \theta_j^2)$. If it is $o_{ij}^1$, then agent 2 receives 1 in this scenario and has an incentive to misreport its type as $\theta_i^2$, since $o(\theta_i^1, \theta_i^2) = o_{ii}^H$, which gives it utility 2. On the other hand, if it is $o_{ij}^2$, then agent 1 receives 1 in this scenario and has an incentive to misreport its type as $\theta_j^1$, since $o(\theta_j^1, \theta_j^2) = o_{jj}^H$, which gives it utility 2. But this contradicts incentive compatibility.

Finally, we show that $S$ has at least $K$ elements. Suppose the agents' types are $\theta_i^1$ and $\theta_j^2$. If $(i,j)$ is an edge, which happens with probability $\frac{2m}{n^2}$, we get a social welfare of $5n^2 + 1$. If $i = j$ and $i \notin S$, which happens with probability $\frac{n-|S|}{n^2}$, we get a social welfare of at most 2. If $i = j$ and $i \in S$, which happens with probability $\frac{|S|}{n^2}$, we get a social welfare of 4. Finally, if $i \neq j$ and $(i,j)$ is not an edge, which happens with probability $\frac{n^2-2m-n}{n^2}$, we get a social welfare of at most 4. Adding up all the terms we find that we get a social welfare of at most $\frac{2m(5n^2+1)+2(n-|S|)+4|S|+4(n^2-2m-n)}{n^2}$. But this must be at least $G = \frac{2m(5n^2+1)+2(n-K)+4K+4(n^2-2m-n)}{n^2}$. It follows that $|S| \geq K$.

∎

We now move on to Bayes-Nash implementation. For this, we will reduce from the $\mathcal{NP}$-complete KNAPSACK problem [8].

**Definition 7 (KNAPSACK)** *We are given a set $I$ of $m$ pairs of (nonnegative) integers $(w_i, v_i)$, a constraint $C > 0$ and a goal $D > 0$. We are asked whether there exists a subset $S \subseteq I$ such that $\sum_{j \in S} w_j \leq C$ and $\sum_{j \in S} v_j \geq D$.*

**Theorem 2** *DETERMINISTIC-MECHANISM-DESIGN with Bayes-Nash implementation (DMDBN) is NP-complete, even with 2 agents and the social welfare function as the objective function.*

**Proof**: To show that the problem is in $\mathcal{NP}$, we observe that specifying an outcome function here requires only the specification of $|\Theta^1| \times |\Theta^2|$ outcomes, and given such an outcome function we can verify whether it meets our requirements in polynomial time. To show that it is $\mathcal{NP}$-hard, we reduce an arbitrary KNAPSACK instance to the following DMDBN instance. Let $W = \sum_{j \in I} w_j$ and $V = \sum_{j \in I} v_j$. There are $m+2$ outcomes: $o_1, o_2, \ldots, o_m, o_{m+1}, o_{m+2}$. We have $\Theta_1 = \{\theta_1^1, \theta_2^1, \ldots, \theta_m^1\}$, where $\theta_j$ occurs with probability $\frac{w_j}{W}$; and $\Theta_2 = \{\theta_1^2, \theta_2^2\}$, where each of these types occurs with probability $\frac{1}{2}$. The utility functions are as follows: for all $j$, $u_1(\theta_j^1, o_j) = (\frac{v_j}{w_j} + 1)W$; $u_1(\theta_j^1, o_{m+2}) = -W$; and $u_1$ is 0 everywhere else. Furthermore, $u_2(\theta_1^2, o_{m+1}) = W$; $u_2(\theta_1^2, o_{m+2}) = W - C$; $u_2(\theta_2^2, o_{m+2}) = W(2V+1)$; and $u_2$ is 0 everywhere else. We set $G = WV + \frac{W+D}{2}$. We now proceed to show that the two problem instances are equivalent.

First suppose there is a solution to the KNAPSACK instance, that is, a subset $S$ of $I$ such that $\sum_{j \in S} w_j \leq C$ and $\sum_{j \in S} v_j \geq D$. Then let the outcome function be as follows: $o(\theta_j^1, \theta_1^2) = o_j$ if $j \in S$, $o(\theta_j^1, \theta_1^2) = o_{m+1}$ otherwise; and for all $j$, $o(\theta_j^1, \theta_2^2) = o_{m+2}$. First we show that truthtelling is a BNE. Clearly, agent 1 never has an incentive to misrepresent its type, since if agent 2 has type $\theta_2^2$, the type that 1 reports makes no difference; and if agent 2 has type $\theta_1^2$, misrepresenting its type will lead to utility at most 0 for agent 1, whereas truthful reporting will give it utility at least 0. It is also clear that agent 2 has no incentive to misrepresent its type when its type is $\theta_2^2$. What if agent 2 has type $\theta_1^2$? Reporting truthfully will give it utility $\sum_{j \in I-S} \frac{w_j}{W} W = \sum_{j \in I-S} w_j \geq W - C$, whereas reporting $\theta_2^2$ instead will give it utility $W - C$. So there is no incentive for agent 2 to misrepresent its type in this case, either. Now, we show that the expected value of $g$ attains the goal. With probability $\frac{1}{2}$ agent 2 has type $\theta_2^2$ and the social welfare is $W(2V+1) - W = 2WV$. On the other hand, if agent 2 has type $\theta_1^2$, the expected social welfare is $\sum_{j \in I-S} \frac{w_j}{W} W + \sum_{j \in S} \frac{w_j}{W}((\frac{v_j}{w_j}+1)W) = W + \sum_{j \in S} v_j \geq W + D$. So, the total expected social welfare is greater or equal to $\frac{2WV+W+D}{2} = G$. Hence, the mechanism design instance has a solution.

Now suppose the mechanism design instance has a solution, that is, a function $o$ satisfying the desired properties. First suppose that there exists a $\theta_j^1$ such that $o(\theta_j^1, \theta_2^2) \neq o_{m+2}$. The social welfare in this case can be at most $(\frac{v_j}{w_j} + 1)W < WV$ (since $w_j$ is a nonnegative integer). Now, the event $(\theta_j^1, \theta_2^2)$ occurs with probability at least $\frac{1}{2W}$ (again, since $w_j$ is a nonnegative integer), and it follows that the probability of an event with social welfare $2WV$ occuring is at most $\frac{W-1}{2W}$. (For this can happen only when agent 2 has type $\theta_2^2$.) It follows that the maximal contribution to the expected social welfare that we can expect from the cases where agent 2 has type $\theta_2^2$ is $\frac{1}{2W}WV + \frac{W-1}{2W}2WV = (W - \frac{1}{2})V$. Additionally, it is easy to see that the maximal contribution to the expected social welfare that we can expect from the cases where agent 2 has type $\theta_1^2$ is $\frac{1}{2} \sum \frac{w_j}{W}(\frac{v_j}{w_j} + 1)W = \frac{1}{2}(W + V)$. It follows that the total expected social welfare can be no more than

$(W - \frac{1}{2})V + \frac{1}{2}(W + V) = WV + \frac{W}{2}$. But this is smaller than $G$, contradicting our assumptions about $o$. It follows that for all $\theta_j^1$, we have $o(\theta_j^1, \theta_2^2) = o_{m+2}$. Now, let $S = \{j | o(\theta_j^1, \theta_1^2) = o_j\}$. We claim $S$ is a solution to the KNAPSACK instance. First, observe that the expected utility that agent 2 gets from misrepresenting its type as $\theta_2^2$ when it is really $\theta_1^2$ is $W - C$. Thus, in order for agent 2 not to have an incentive to do this, we must have $\sum\limits_{j | o(\theta_j^1, \theta_1^2) = o_{m+1}} \frac{w_j}{W} W \geq W - C$. But the left-hand side of the inequality is smaller or equal to $\sum\limits_{j \in I-S} w_j$. It follows that $\sum\limits_{j \in I-S} w_j \geq W - C$, or $\sum\limits_{j \in S} w_j \leq C$. Second, the expected social welfare can be expressed as $\frac{2WV}{2} + \frac{1}{2} \sum\limits_{j \in I-S} \frac{w_j}{W} W + \frac{1}{2} \sum\limits_{j \in S} \frac{w_j}{W}(\frac{v_j}{w_j} + 1)W = WV + \frac{W}{2} + \frac{1}{2} \sum\limits_{j \in S} v_j$. But by our assumptions on the outcome function, we know that this must be greater or equal to $G = WV + \frac{W+D}{2}$. It follows that $\sum\limits_{j \in S} v_j \geq D$. So we have found a solution to the KNAPSACK instance. ∎

## 4 Randomized mechanisms

Randomized mechanisms are a generalization of deterministic mechanisms, and as such potentially allow one to increase the expectation of the (social) objective function. In this section we show that randomized mechanisms also allow one to circumvent the complexity problems of deterministic mechanism design.

**Definition 8** *Given a preference aggregation setting, a* randomized mechanism *is a function that, given any vector of reported types, produces a probability distribution over the outcome set. That is, it is a function $p : \Theta^1 \times \Theta^2 \times \ldots \times \Theta^N \to \mathcal{PROBDISTS}(O)$.*

We need to make only minor modifications to our definitions of solution concepts and the computational problem of mechanism design to accommodate for this generalization. In these definitions, $E_{o \leftarrow p}$ means that the expectation is taken when $o$ is randomly chosen according to the distribution $p$.

**Definition 9** *Given a preference aggregation setting, a mechanism is said to* implement its outcome function in dominant strategies *if truthtelling is always optimal even when the types reported by the other agents are already known. For a randomized mechanism, this means that for any $i \in A$, for any type vector $(\theta^1, \theta^2, \ldots, \theta^i, \ldots, \theta^N) \in \Theta^1 \times \Theta^2 \times \ldots \times \Theta^i \times \ldots \times \Theta^N$, and for any $\theta_0^i \in \Theta^i$, we have $E_{o \leftarrow p(\theta^1, \theta^2, \ldots, \theta^i, \ldots, \theta^N)}(u_i(\theta^i, o)) \geq E_{o \leftarrow p(\theta^1, \theta^2, \ldots, \theta_0^i, \ldots, \theta^N)}(u_i(\theta^i, o))$*

**Definition 10** *Given a preference aggregation setting, a mechanism is said to* implement its outcome function in Bayes-Nash equilibrium *if truthtelling is always optimal as long as the other agents' types are unknown, and the other agents report truthfully. For a randomized mechanism, this means that for any $i \in A$, and for any $\theta^i, \theta_0^i \in \Theta^i$, we have $E_{\theta^{-i}}(Eo \leftarrow p(\theta^1, \theta^2, \ldots, \theta^i, \ldots, \theta^N)(u_i(\theta^i, o))) \geq E_{\theta^{-i}}(Eo \leftarrow p(\theta^1, \theta^2, \ldots, \theta_0^i, \ldots, \theta^N)(u_i(\theta^i, o)))$.*

**Definition 11 (RANDOMIZED-MECHANISM-DESIGN)** *We are given a preference aggregation setting, a solution concept, and an objective function $g : \Theta^1 \times \Theta^2 \times \ldots \times \Theta^n \times O \to \mathcal{Q}$ with a goal $G$. We are asked whether there exists a randomized mechanism for the preference aggregation setting which*

- *satisfies the given solution concept, and*
- *attains the goal, i.e. $E_{\theta^1, \theta^2, \ldots, \theta^N}(E_{o \leftarrow p(\theta^1, \theta^2, \ldots, \theta^N)}(g(\theta^1, \theta^2, \ldots, \theta^N, o))) \geq G$.*

We now demonstrate how randomization in the mechanism allows us to compute optimal mechanisms much easier: we merely need to formulate the RANDOMIZED-MECHANISM-DESIGN instance as a linear program. For ease of exposition, we demonstrate how to do this only in the two-agent case. However, the method readily generalizes to larger numbers of agents: the theorems below hold for any constant number of agents. (As we indicated before, the description length of $o$ is exponential in the number of agents, so this approach is tractable only for small numbers of agents.)

**Theorem 3** *2-agent RANDOMIZED-MECHANISM-DESIGN with dominant strategies implementation as the solution concept is solvable in polynomial time by linear programming, for any (polynomially computable) objective function.*

**Proof**: Let $p_{ij}^k$ denote the probability of choosing $o_k$ when the reported types are $\theta_i^1$ and $\theta_j^2$. That is, $p_{ij}^k = (p(\theta_i^1, \theta_j^2))(o_k)$. These will be the variables that the linear program is to determine. Note there are polynomially many of them ($|\Theta^1||\Theta^2||O|$). We introduce the following constraints in the linear program (corresponding to the requirements of implementation in dominant strategies):

- For every $\theta_j^2 \in \Theta^2$, for every $\theta_i^1, \theta_l^1 \in \Theta^1$, we have: $\sum\limits_{k: o_k \in O} p_{ij}^k u_1(\theta_i^1, o_k) \geq \sum\limits_{k: o_k \in O} p_{lj}^k u_1(\theta_i^1, o_k)$;

- For every $\theta_i^1 \in \Theta^1$, for every $\theta_j^2, \theta_l^2 \in \Theta^2$, we have: $\sum\limits_{k: o_k \in O} p_{ij}^k u_2(\theta_j^2, o_k) \geq \sum\limits_{k: o_k \in O} p_{il}^k u_2(\theta_j^2, o_k)$.

Finally, we seek to maximize the following expression (which is just the expectation of the objective function):

- $\sum_{i:\theta_i^1 \in \Theta^1} \sum_{j:\theta_j^2 \in \Theta^2} \sum_{k:o_k \in O} p_1(\theta_i^1) p_2(\theta_j^2) p_{ij}^k g(\theta_i^1, \theta_j^2, o_k).$

Note that all the expressions are indeed linear in the $p_{ij}^k$, and there is a polynomial number of inequalities $(|\Theta^1|^2|\Theta^2| + |\Theta^1||\Theta^2|^2)$. ∎

**Theorem 4** *2-agent RANDOMIZED-MECHANISM-DESIGN with Bayes-Nash implementation as the solution concept is solvable in polynomial time by linear programming, for any (polynomially computable) objective function.*

**Proof**: The $p_{ij}^k$ are as before. We introduce the following constraints in the linear program (corresponding to the requirements of implementation in Bayes-Nash equilibirum):

- For every $\theta_i^1, \theta_l^1 \in \Theta^1$, we have:
  $\sum_{j:\theta_j^2 \in \Theta^2} \sum_{k:o_k \in O} p_2(\theta_j^2) p_{ij}^k u_1(\theta_i^1, o_k) \geq$
  $\sum_{j:\theta_j^2 \in \Theta^2} \sum_{k:o_k \in O} p_2(\theta_j^2) p_{lj}^k u_1(\theta_i^1, o_k);$

- For every $\theta_j^2, \theta_l^2 \in \Theta^2$, we have:
  $\sum_{i:\theta_i^1 \in \Theta^1} \sum_{k:o_k \in O} p_1(\theta_i^1) p_{ij}^k u_2(\theta_j^2, o_k) \geq$
  $\sum_{i:\theta_i^1 \in \Theta^1} \sum_{k:o_k \in O} p_1(\theta_i^1) p_{il}^k u_2(\theta_j^2, o_k).$

Again, we seek to maximize the following expression:

- $\sum_{i:\theta_i^1 \in \Theta^1} \sum_{j:\theta_j^2 \in \Theta^2} \sum_{k:o_k \in O} p_1(\theta_i^1) p_2(\theta_j^2) p_{ij}^k g(\theta_i^1, \theta_j^2, o_k).$

Note that all the expressions are indeed linear in the $p_{ij}^k$, and there is a polynomial number of inequalities $(|\Theta^1|^2 + |\Theta^2|^2)$. ∎

## 5 Increasing economic efficiency through randomization—a connection to computational complexity

It is known in game theory that allowing for randomization in the mechanism can increase expected social welfare (or other objective functions). Interestingly, our results from above would prove this fact *through connections to computational complexity*, if $\mathcal{P} \neq \mathcal{NP}$. For suppose it were never possible to increase social welfare via randomization. Then, since randomized mechanisms are a generalization of deterministic mechanisms, it would follow that the expected social welfare from the best randomized mechanism would equal the expected social welfare from the best deterministic mechanism. Therefore, to solve a DETERMINISTIC-MECHANISM-DESIGN instance, we could simply solve the corresponding RANDOMIZED-MECHANISM-DESIGN instance. But we have shown that for some solution concepts, DETERMINISTIC-MECHANISM-DESIGN is $\mathcal{N}P$-complete while RANDOMIZED-MECHANISM-DESIGN is in $\mathcal{P}$. So, we could conclude that $\mathcal{P} = \mathcal{NP}$.

The fact that randomization in the mechanism can increase expected social welfare is also easy to prove directly, as the following example shows. Let there be 3 outcomes: $o_1$, $o_2$, and $o_3$. Agent 1 has 2 types (both equally likely), $\theta_1^1$ and $\theta_2^1$. Agent 2 only has 1 type, $\theta_1^2$. The utility functions are as follows: for agent 1, $u_1(\theta_1^1, o_1) = 1$; $u_1(\theta_1^1, o_2) = 2$; $u_1(\theta_1^1, o_3) = 0$; $u_1(\theta_2^1, o_1) = 8$; $u_1(\theta_2^1, o_2) = 2$; and $u_1(\theta_2^1, o_3) = 0$. For agent 2, $u_2(\theta_1^2, o_1) = 0$; $u_2(\theta_1^2, o_2) = 0$; and $u_2(\theta_1^2, o_3) = 4$. Because there is only one agent who has more than one type, implementation in dominant strategies and in Bayes-Nash equilibrium (and indeed all reasonable solution concepts) coincide. It is easy to show that the deterministic mechanism that maximizes social welfare among the nonmanipulable ones is given by $o(\theta_1^1, \theta_1^2) = o_2$; $o(\theta 2^1, \theta_1^2) = o_1$, for an expected social welfare of 5. (Every mechanism that does not choose $o_1$ when agent 1 has type $\theta_2^1$ will have expected social welfare no greater than 4. If the mechanism does choose $o_1$ in this case, we cannot choose $o_3$ in the case where agent 1 has type $\theta_1^1$, because agent 1 would have an incentive to misreport type $\theta_2^1$ when its true type is $\theta_1^1$. So, the best the mechanism can do is to select $o_2$ in this case.) However, if we allow for randomization even just in the case where agent 1 has type $\theta_1^1$, we can do better by choosing $o_2$ with probability $\frac{1}{2}$, and $o_3$ with probability $\frac{1}{2}$. (This gives an expected social welfare of $5\frac{1}{2}$, and there is no incentive for agent 1 to manipulate because the expected utility that it gets from reporting truthfully in the case where its type is $\theta_1^1$ is 1—which is the same as it would get by misreporting $\theta_2^1$.)

## 6 Related research

While we are, to our knowledge, the first to study the computational complexity of mechanism design, there has been a significant amount of research on other aspects of computing in games. For example, there has been considerable work on finding equilibria in games. AI work on this topic has focused on novel knowledge representations which, in certain settings, can drastically speed up equilibrium finding (e.g. [11–13]).

A recent stream studies the complexity of *executing* mechanisms: voting mechanisms [4], combinatorial auctions (e.g. [18]), and other optimal and approximate mechanisms (e.g. [17]).

Another research stream has focused on determining the computational complexity of *manipulating* mechanisms, with the goal of designing mechanisms where constructing a beneficial manipulation is hard [2, 3, 6].

Yet another stream has focused on games where the agents need to compute their preferences, and have limited computing available. In that setting, computational complexity actually affects the equilibrium, not merely the complexity of finding one [14, 15].

## 7 Conclusions and future research

The aggregation of conflicting preferences is a central problem in multiagent systems, be the agents human or artificial. The key difficulty is that the coordinator, who tries to aggregate the preferences, is uncertain about the agents' preferences *a priori*, and the agents may report their preferences insincerely.

Mechanism design is the art of designing the rules of the game so that the agents are motivated to report their preferences truthfully and a (socially) desirable outcome is chosen. We proposed an approach where a mechanism is automatically created for the preference aggregation setting at hand. This approach can be used even in settings that do not satisfy the assumptions of general classical mechanisms. It may also yield better mechanisms (in terms of stronger nonmanipulability guarantees and/or better social outcomes) than the classical mechanisms. Finally, it may allow one to circumvent impossibility results (such as the Gibbard-Satterthwaite theorem) which state that there is no mechanism that is desirable across all preferences.

The downside is that the mechanism design optimization problem needs to be solved anew each time. Focusing on settings where side payments are not possible, we showed that the mechanism design problem is $\mathcal{NP}$-complete for deterministic mechanisms. This holds both for dominant-strategy implementation and for Bayes-Nash implementation. We then showed that if we allow randomized mechanisms, the mechanism design problem becomes solvable in polynomial time in both cases. In other words, the coordinator can tackle the computational complexity introduced by its uncertainty about the agents' preferences by making the agents face additional uncertainty. This comes at no loss, and in some cases at a gain, in the (social) objective.

Future research includes extending the approach of automated mechanism design to settings where side payments are viable but the classical general mechanisms do not work (for example because the agents do not have quasilinear preferences or additional requirements are posed, such as strong budget balance). Another interesting use of automated mechanism design is to solve for mechanisms for a variety of settings (real or artificially generated), and to see whether general mechanisms (or mechanism design principles) can be inferred.